\documentstyle[prb,twocolumn,aps]{revtex}

\tighten
\begin{document}
\title{Non-Linear Effects in Resonant Tunneling; Bistabilities and Self-Sustained
Oscillating Currents}
\author{E. S. Rodrigues$^1$, E. V. Anda$^2$, P.Orellana$^3$ and F. Claro$^4$}
\address{$1$ Instituto de F\'{\i}sica, Universidade Federal\\
Fluminense,\\
Av. Litor\^anea s/n$^{\underline o}$, Gragoat\'a, Niter\'oi RJ,\\
Brasil, 24210-340.\\
$2$ Departamento de F\'{\i}sica, Pontif\'{\i}cia\\
Universidade Cat\'olica do Rio de Janeiro, \\
Caixa Postal 38071, Rio de Janeiro, Brasil, 22452-970.\\
$3$ Departamento de F\'{\i}sica, Universidad Cat\'olica\\
del Norte,\\
Av. Angamos 0610, Casilla 1280, Antofagasta, Chile.\\
$4$ Facultad de F\'{\i}sica, Pontificia Universidad\\
Cat\'olica del Chile,\\
Av. Vicu\~na Mackenna 4860, Casilla 306 Stgo 22}
\maketitle

\begin{abstract}
We study non-linear phenomena in double barrier heterostructures. Systems in
3D under the effect of an external magnetic field along the current and 1D
systems are analyzed. Non-linearities are reflected in the I-V
characteristic curve as bistabilities, instabilities and time dependent
oscillations of the currents. The nature of the non-linear behavior depends
upon the parameters that define the system.
\end{abstract}

\makeatletter

\global\@specialpagefalse

\let\@evenhead\@oddhead
\makeatother

\narrowtext

\section{Introduction}

Resonant tunneling in double barrier heterostructures (DBH) have been found
to exhibit a variety of interesting physical properties \cite{ll}. Transport
in a 3D DBH presents non-linear phenomena, which are reflected in the
observation of multistabilities in the I-V characteristic curve \cite{vj}.
This property is produced by the rapid leakage of the electronic charge
accumulated at the well between the barriers when the applied potential is
just taking the device out of resonance. It is a result of the Coulomb
charge interaction at the well and has been theoretically studied, in a 3D
system, using a mean field approximation \cite{po}. In this work we have
found a similar behavior in a 1D system constituted by a quantum dot
connected to two leads. Although, the origin of the bistability is the
non-linearity produced by the Coulomb interaction, as in the 3D system, what
is dominant in a quantum dot is the spatial e-e correlation. It produces
simultaneously the Coulomb blockade of the charges, that appears as periodic
oscillations of the conductance as a function of the gate potential and the
bistability of the current. For a 3D system instead, the Coulomb blockade is
absent and the electronic interaction can be incorporated as a
renormalization of the quasi-particle energies using a mean field
approximation which gives rise to the non-linear effects.

Non-linear time dependent transport in double-barrier structures is another
very interesting phenomena. In this work we report the presence of
self-sustained current oscillations when a large magnetic field is applied
along the direction of the current in a 3D DBH. The magnetic field produces
a drastic change in the number of resonant electrons going through the well
inducing a great enhancement of the non linearities of the system. The state
of charge of the well controls the oscillatory behavior of the current. Our
results show that the system bifurcates as the magnetic field is increased
and may transit to chaos at large enough field. The oscillations are
exponentially sensible to initial conditions or to infinitesimal changes in
the parameter of the Hamiltonian, characterizing a chaotic behavior which
has been studied in detail in the parameter space.

We obtain a similar result for the current going along the quantum dot
connected to two leads. The produced self-sustained time dependent current
oscillations is controlled in this case by the voltage applied to the
metallic gate. In both cases the period of the oscillations depends upon the
transit time of the electron going along the well, which is determined by
the size of the barriers.

\section{The Model}

Consider a 1D or a 3D double barrier structure (DBS) device. In order to
study its time evolution under a bias we adopt a first-neighbors
tight-binding model for the Hamiltonian. For the 3D situation we include an
external magnetic field B in the growth direction. We assume a parabolic
energy dispersion parallel to the interfaces while the field quantizes the
motion of the electrons in the perpendicular plane giving rise to Landau
levels with energies $\epsilon_n = (n + \frac{1}{2}) \hbar \omega_c$, where $%
n = 0, 1, 2, ...$ is the Landau index and $\omega_c = eB / m^*c$ is the
cyclotron frequency. The probability amplitude $b_j^{nk}$ for an electron in
a time dependent state $\vert kn>$ at plane $j$ along $z$ in the Landau
level $n$ is determined by the equation of motion

\begin{eqnarray}
i\hbar \frac{db_{j}^{nk}}{dt} &=&\biggl( \epsilon _{j}+\epsilon
_{n}+U\sum_{mk^{\prime }}|b_{j}^{mk^{\prime }}|^{2}\biggr) b_{j}^{nk} 
\nonumber \\
&&+v\biggl( b_{j-1}^{nk}+b_{j+1}^{nk}-2b_{j}^{nk}\biggr),
\end{eqnarray}

\noindent where $v$ is the hopping matrix elements between electrons in
nearest-neighbor planes, and $\epsilon_j$ represents the band contour and
external bias. Here the sum over $(m,k)$ covers all occupied electron states
of the system, within the various Landau levels $m$ with energies below the
Fermi level, incident on the DBS from the emitter side. In writing equation [%
\ref{mo}] we have adopted a Hartree model for the electron-electron
interaction. We have neglected inter-Landau level transitions since they are
of little importance at not too low magnetic fields \cite{dy}, and averaged
over allowed transitions between degenerate states within a Landau level,
taking advantage of the localization induced by the Gaussian factor in the
Landau basis.

For the 1D system we can write a similar equation where the subindex $n$
corresponding to the Landau level is eliminated. The electron-electron
interaction restricted to the dot has to be treated in this case using a
strong interaction limit approximation capable of including simultaneously
the Coulomb blockade effect and the non-linear behavior. The
electron-electron interaction is analyzed within the context of the Hubbard
I approximation \cite{az}. In this case the equation of motion is written as,

\[
i \hbar \frac{db_j^{k}}{dt} = \epsilon_jb_j^k + v_jb_{-1}^k +
v_{j+1}b_{j+1}^k \hskip 0.4cm (i \ne 0) \mbox{;} 
\]
\begin{equation}
i \hbar \frac{db_o^{k \alpha}}{dt} = \biggl(\epsilon_o^{\alpha} + V_p %
\biggr) b_j^{k \alpha} + n^{\alpha} v_o \biggl(b_{-1}^k + b_1^k \biggr)
\label{ii}
\end{equation}

\noindent where $\alpha = \pm$; $n^{+} = n$; $n^{-} = 1 - n$; $\epsilon^{+}
= U$; $\epsilon^{-} = 0$; $v_j = v \hskip 0.3cm \forall j \ne 0,1$; $v_j =
v_o = 0, 1$; $n = \sum_{k^{\prime}} \vert b_o^{k^{\prime}} \vert ^2$; $b_o^k
= \sum_{\alpha = 1}^2 b_o^{k \alpha}$; $V_p$ is the gate potential applied
at the well and the sum over $k^{\prime}$ covers as before all occupied
electron states of the system.

The time dependent equations are solved using a half-implicit numerical
method which is second-order accurate and unitary \cite{az}.

The approach taken here assumes that in the presence of a bias, the wave
function at time $t$ outside the structure is given by \cite{az}

\begin{eqnarray}
b_{j}^{nk} &=&Ie^{ikr_{j}}+R_{jn}e^{-ikr_{j}}\hskip 0.3cmr\le -L\hskip 1.7cm
\nonumber \\
b_{j}^{nk^{\prime }} &=&T_{jn}e^{ik^{\prime }r_{j}}\hskip 0.3cmr\ge -L
\end{eqnarray}

\noindent Here $k$ and $k^{\prime} = \sqrt{2m^*[\epsilon^k + \epsilon_L]}/
\hbar$ are the wave numbers of the incoming and outgoing states,
respectively, with $\epsilon^k = \epsilon_n - 4vsin^2(\frac{ka}{2})$ the
energy of the incoming particle (for the 1D case $\epsilon_n = 0$) and $%
\epsilon_L$ is the external bias. To model the interaction with the particle
reservoir outside the structure the incident amplitude $I$ is assumed to be
a constant independent of the coordinates. The envelope function of the
reflected and transmitted waves, $R_{jn}$ and $T_{jn}$, are allowed to vary
with $j$, however. Since far from the barriers these quantities are a weak
function of the coordinate $z_j$ we restrict ourselves to the linear
corrections only. This approximations is appropriate provided the time step $%
\delta t$ does not exceed a certain limiting value. For the results
presented here, a maximum value of $\delta t = 3 \times 10^{-17}$s was found
sufficient to eliminate spurious reflections at the boundary while
maintaining numerical stability up to $20 \times 10^{-12}$s.

In our numerical procedure the coefficients obtained for one bias are used
as the starting point for the next bias step.

We study first the 3D DBH under the effect of an external magnetic field.
The model is applied to an asymmetric double barrier structure. The second
barrier is made wider than the first in order to enhance the trapping charge
in the well. The conduction band offset is set at $300$meV. In equilibrium
and at $B = 0T$, the Fermi level lies $19.2$meV above the asymptotic
conduction band edge, so that the zero bias resonance lies well above the
Fermi sea. The parameter values in equation [\ref{mo}] are set at $v = -2.16$%
eV and $U = 100$meV. The latter was chosen phenomenologically so as to fit
the experimental I-V characteristic for a GaAs device in the absence of an
external magnetic field \cite{az}.

For small values of the magnetic field a stationary solution is reached
after a short transient. At large enough bias two stationary solutions
emerge, obtained by either increasing or decreasing the applied voltage.
This corresponds to the well known bistability of 3D DBH.

A completely novel feature starts up as the field is increased. At small
values of the external bias a stationary solution is rapidly reached. As the
bias is increased, however, a critical value arises beyond which no
stationary solution exists, and the system begins to oscillate in a
self-sustained way. After a range of voltages over which the oscillations
persist a stationary solution is reached again.

In fig.1 we show the current at the center of the well for $B = 13$T and a
fixed bias $V = 0.27$v. A transient period is followed by an oscillation
that is never damped out. Although not perfectly periodic, the oscillation
has two strong Fourier components at frequencies $v \sim 0.3$ and $0.8$THz.

For a 1D system we study a model which consists of leads connected to a
quantum dot represented by $v_o = 0.1v$, $U=1.0v$ and a Fermi level $E_f =
0.01v$. The normalization of the wave function is taken so that each site
could have up to a maximum of two electrons. We solve first equations [\ref
{ii}] supposing that they possess stationary solutions. Since the electron
density has rather long-range oscillations we made sure the sample was long
enough to make finite size effects negligible. This is guaranteed taken a
sample of $N>100$ sites. Fig 2a shows the current voltage (I-V)
characteristics for several a gate potential applied at the quantum dot $V_p
= 0.1v$. The system has a bistable behavior, which corresponds to the well
known two solutions for the current; as the voltage is increased when the
dot is charged (solid line), and when the voltage is decreased the solution
when the dot is without charge (dashed line). In either case the
self-consistent solution converges to a stable fixed point after some finite
number of iterations. Similarly to 3D tunneling devices in this 1D structure
an increase of the Fermi level augments the charge content of the well and
consequently the non-linearities and in particular the width of the
hysteresis loop. When the gate potential is greater than a threshold value,
which depends upon the parameters of the system, a completely novel feature
starts to develop where two periods, two fixed points are encountered for
certain voltages as shown in Fig 2b. This bubble-like solution is obtained
as the voltage is increased outside the region of bistability showing that
although both phenomena are derived from the non-linearities introduced by
the local Coulomb repulsion, they corresponds to two different effects. Note
that these new solutions are not reached, as the voltage is decreased
(dashed line) as it corresponds to a discharged dot where the
non-linearities are negligible. As the potential gate is raised still
further, the area of the bistable bubble increases and it enters into de
bistable region. At larger gate potentials the solution undergoes further
bifurcations and finally a bifurcation cascade leading to a chaotic region.
These behavior is in fact indicating the absence of stationary solution of
the system.

Similarly to the 3D DBH under the effect of a magnetic field, self-sustained
oscillations for the current are found in the case of a dot connected to two
leads in the parameter region where there is no stationary solutions.

We interpret these results for both systems in the following way. In
general, current flows through the heterostructure as long as a tunneling
resonance lies within the emitter Fermi sea. We assume that at very low bias
the resonance lies above the Fermi energy so that no current flows. As the
bias increases it drops, reaching eventually the Fermi energy, thus opening
a channel for electrons to tunnel through the double barrier. A current is
thus established. The charge is trapped in the potential well, raising its
bottom and pushing the resonance towards higher energies for the 3D system.
For the 1D system the weight of the resonance level is proportional to $1 -
n $ and as a consequence reduces when the charge increases. The current
drops, some of the accumulated charge leaks out allowing the current to flow
more easily once again, and a new cycle begins.

According to the picture drawn above the charge in the well lags the
current, as exhibited in the inset of fig.1. Here the charge at the center
of the well (dashed line) is plotted together with the current at the same
point (full line), the former displaced a time $\tau \sim 1.4$ps to the left
with respect to the latter. The relaxation time $\tau$ and the period of the
oscillations are determined by the tunneling time for the electrons to leak
out through the barriers and may be adjusted by modifying the barrier
thickness.

\section{Conclusions}

There is a wide region of accessible values of these parameters were
bistability, instabilities and oscillations in the current are either
present or absent. This opens up the interesting possibility of
experimentally studying the transition from one behavior to the other as the
parameters of both systems studied are varied. From the view point of
possible applications these systems could operate as electromagnetic
generators in the THz regime.

\section{Acknowledgments}

This work was supported by FONDECYT Grants. 1980225 and 1960417, and
DGICT-UCN (Chile), CNPq and FINEP (Brazil).

\begin{figure}[t]
\hbox{{\vbox{\input epsf
\epsfxsize 1.0\hsize\epsfbox{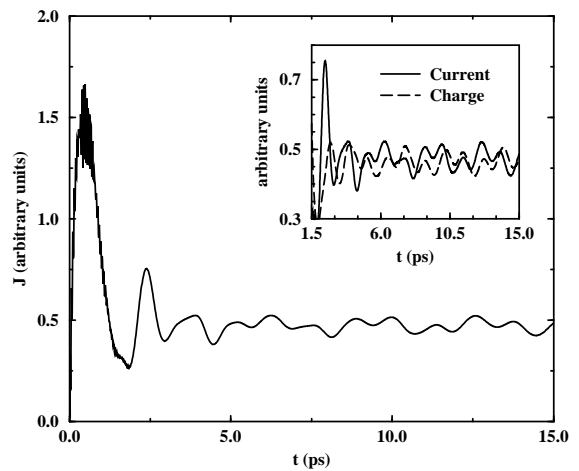}}}}
\caption{Current as a function of time at the center of the well for a 3D
DBH under the effect of an applied magnetic field 13T. The inset shows the
charge and the current shifted one relative to the other by $1.4 ps$.}
\label{fig1}
\end{figure}

\begin{figure}[b]
\hbox{{\vbox{\input epsf
\epsfxsize 1.0\hsize\epsfbox{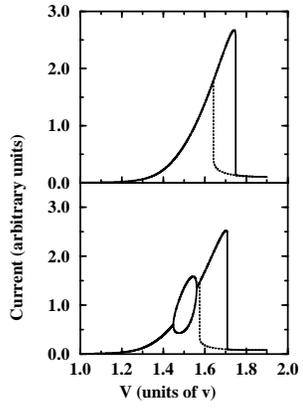}}}}
\caption{I-V characteristic of a quantum dot connected to 1D leads
for:\break a) $V_p = 0.1v$ and b) $V_p = 0.2v$}
\label{fig2}
\end{figure}

\end{document}